\RequirePackage{ifpdf}
\ifpdf 
\documentclass[pdftex]{sigma}
\else
\documentclass{sigma}
\fi

\begin{document}

\allowdisplaybreaks

\renewcommand{\PaperNumber}{105}

\FirstPageHeading

\ShortArticleName{Dolbeault Complex on $S^4\backslash \{\cdot\}$ and $S^6\backslash\{\cdot\}$ through Supersymmetric Glasses}

\ArticleName{Dolbeault Complex on $\boldsymbol{S^4\backslash \{\cdot\}}$ and $\boldsymbol{S^6\backslash\{\cdot\}}$\\ through Supersymmetric Glasses}

\Author{Andrei V.~SMILGA}

\AuthorNameForHeading{A.V.~Smilga}

\Address{SUBATECH, Universit\'e de Nantes,  4 rue Alfred Kastler, BP 20722, Nantes 44307, France\footnote {On leave of absence from ITEP, Moscow, Russia.}}
\Email{\href{mailto:smilga@subatech.in2p3.fr}{smilga@subatech.in2p3.fr}}

\ArticleDates{Received June 22, 2011, in f\/inal form November 09, 2011;  Published online November 15, 2011}

\Abstract{$S^4$ is not a complex manifold, but it is suf\/f\/icient to remove one point to make
it complex. Using supersymmetry methods, we show that the Dolbeault complex
(involving the holomorphic exterior derivative $\partial$ and its Hermitian
conjugate) can be perfectly well def\/ined in this case. We calculate
the spectrum of the Dolbeault Laplacian. It involves 3 bosonic zero modes
such that the Dolbeault index on $S^4\backslash \{\cdot\} $ is equal to 3.}


\Classification{32C15; 53B35; 53Z05}

\section{Introduction}

We start with reminding the standard def\/inition for a complex manifold.
 Suppose a manifold of dimension $D = 2d$ is
covered by several overlapping $D$-dimensional disks. Suppose that in
each such map complex coordinates $w^{j = 1,\ldots,d}$,
$\bar w^{\bar j = 1,\ldots,d} $ are introduced such that
the metric has a Hermitian form
 \begin{gather*}
ds^2 \ =\ h_{j\bar k}(w, \bar w) dw^j d \bar w^{\bar k}   , \qquad
h_{j\bar k}^* = h_{k\bar j}  .
 \end{gather*}
In the region where
a couple of the maps with coordinates $w$, $\bar w$ and $\tilde w$,
$\tilde {\bar w}$ overlap the latter are expressed into one another.
The manifold is called {\it complex} if this relationship
can be made holomorphic, $\tilde w^j = f^j(w^k)$.

For example, $S^2$ (actually, any 2-dimensional manifold) is complex.
To see that, introduce the stereographic complex coordinates
\[
w = \frac {x + iy} {\sqrt{2}} , \qquad  \bar w = \frac {x - iy} {\sqrt{2}}
\]
 such that
 \begin{gather*}
ds^2   =  \frac {2 dw d \bar w}{(1 + \bar w w)^2}  .
  \end{gather*}
This map covers the whole sphere except its north pole (corresponding to $w = \infty$).
Introduce now another stereographic map that covers the whole sphere but its south pole.
The metric is again
 \begin{gather*}
ds^2   =  \frac {2 d \tilde w d {\tilde {\bar w}}}{(1 + {\tilde {\bar w}} \tilde w)^2}  .
  \end{gather*}
 In the region where the maps overlap (the whole sphere but two points), the holomorphic
relation $\tilde w = 1/w$ holds.

Let us try to do the same for $S^4$. Again, we can cover it by two stereographic maps with the coordinates $w_j$ and
$\tilde w_j$\,\footnote{$j = 1,2$ and, when going down to $S^4$ with its conformally
f\/lat metric, we will  not bother to distingush between
covariant and contravariant indices. Neither will we distiguish in this case
the indices $j$ and $\bar j$. The summation over the repeated indices in equations (\ref{metrS4}), (\ref{metrS4tilde})
 and in all the formulas in Sections~\ref{section2}--\ref{section4} is assumed, as usual.}
such that the metric is, on one hand,
 \begin{gather}
\label{metrS4}
ds^2   =  \frac {2 d  w_j d  \bar w_j}{(1 +  \bar w w)^2}  .
  \end{gather}
($\bar w w \equiv \bar w_j w_j = (x^2 + y^2 + z^2 + t^2)/2$) and, on the other hand,
\begin{gather}
\label{metrS4tilde}
ds^2   =  \frac {2 d \tilde w_j d {\tilde {\bar w}}_j}{(1 + {\tilde {\bar w}} \tilde w)^2} .
  \end{gather}
But the relationship $\tilde w_j = \bar w_j/(\bar w w)$ {\it is} not holomorphic anymore meaning that $S^4$ is not
complex.

For complex compact manifolds, one can consider a set of holomorphic $(p,0)$-forms, introduce the operator of exterior holomorphic
derivative $\partial$, its Hermitian conjugate $\partial^\dagger$ and def\/ine  thereby the {\it Dolbeault} complex~(see e.g.~\cite{EGH}).
The operators $\partial$ and $\partial^\dagger$ are nilpotent and the Hermitian
Dolbeault Laplacian $\partial  \partial^\dagger +
\partial^\dagger \partial$ commutes with both $\partial$ and $\partial^\dagger$.
This algebra is isomorphic to the simplest supersymmetry algebra,
 \[
Q^2   =  \bar Q^2 =   0, \qquad \{Q, \bar Q\} = H   .
 \]
 The supersymmetric description of the Dolbeault complex for any (not necessarily K\"ahler) complex manifold
has been constructed in recent~\cite{IvSm}. The superf\/ield action
(f\/irst written in~\cite{Hull})
 is expressed in terms of $d+d$ chiral and antichiral
superf\/ields
\begin{gather*}
W^j   =  w^j + \sqrt{2} \theta \psi^j - i\theta \bar \theta \dot{w}^j, \qquad
\bar W^{\bar j}   =   \bar w^{\bar j} - \sqrt{2} \bar\theta   \bar\psi^{\bar j} + i\theta \bar \theta   \dot{\bar w}^{\bar j}  ,
\\
S = \int dt d^2\theta \left[ - \frac 14 h_{j\bar k} \big(W^l, \bar W^{\bar l}\big)
 {DW^j \bar D \bar W^{\bar k}} + G(\bar W, W) \right]  .
 \end{gather*}
Deriving with this action the component Lagrangian,
then classical and quantum Hamiltonian, using the N\"other theorem, and accurately
resolving the ordering ambiguities~\cite{howto}, we arrive at the expressions for the quantum supercharges
  \begin{gather}
  Q = \psi^c e^k_c\left[\Pi_k -\frac{i}{4} \partial_k  (\ln \det h)
+ i \psi{\,}^b \bar\psi{\,}^{\bar a} \Omega_{k, \bar a b}\right], \nonumber\\
  \bar Q = \bar\psi{\,}^{\bar c} e^{\bar k}_{\bar c}
\left[\bar\Pi_{\bar k} -\frac{i}{4} \partial_{\bar k} (\ln \det h)
+ i \bar\psi{\,}^{\bar b} \psi{\,}^{a} \bar{\Omega}_{\bar k, a \bar b}\right] , \label{Qcovgen}
\end{gather}
where $e_j^c$ are the vielbeins, $e_j^c \bar e_{\bar k}^{\bar c} = h_{j\bar k}$,
chosen such that $\det e = \det \bar e   =  \sqrt{\det h}$,
\[
\Omega_{j, \bar  b a} \equiv \Omega_{j, \ a}^{\ \ b}   = e^b_p \big(\partial_j e^p_a + \Gamma^p_{jk} e^k_a\big)
\]
(and the complex conjugate $\bar \Omega_{\bar j, b \bar a} \equiv \Omega_{\bar j, \  \bar a}^{\ \ \bar b} $)
are the holomorphic and antiholomorphic components of the standard Levi-Civita spin connections\footnote{In the K\"ahler case, nonholomorphic components like $\Omega_{j \  \bar b}^{\ a}$ vanish. For  generic
complex manifold, they do not vanish (though they vanish again for some special torsionful connections (\ref{hatOm})
below)  but {\it do} not enter the supercharges~(\ref{Qcovgen}). We refer the reader to~\cite{IvSm} and
to recent~\cite{HRR} for further pedagogical explanations.},
$\bar \psi^{\bar a}  = \partial/\partial \psi^a$, and
  \begin{gather*}
\Pi_k = -i\left(\frac{\partial}{\partial z^k} -  \partial_k G\right), \qquad
\bar\Pi_{\bar k} = -i \left(\frac{\partial}{\partial \bar z{\,}^{\bar k}} +  \partial_{\bar k} G\right).
\end{gather*}
are the covariant derivatives involving the gauge f\/ield
\begin{gather}
\label{A}
A_{j,\bar k} = (-i\partial_j G, i\bar \partial_{\bar k} G)   .
 \end{gather}

 The quantum supercharges (\ref{Qcovgen}) act on the wave
functions
   \begin{gather*}
\Psi\big(w^j, \bar w^{\bar k}; \psi^a\big)   = A^{(0)}\big(w^j, \bar w^{\bar k}\big) + \psi^a A^{(1)}_a\big(w^j, \bar w^{\bar k}\big) + \cdots
+ \psi^{a_1}\cdots \psi^{a_d}A^{(d)}_{[a_1 \cdots a_d]}\big(w^j, \bar w^{\bar k}\big).
 \end{gather*}
 The components of this wave function $A^{(0)}$, $A^{(1)}_a $, etc. can be mapped onto the space of
the holomorphic forms $A^{(0)}$,  $e^a_j A^{(1)}_a   dw^j$, etc. A $(p,0)$-form corresponds to the wave
function with the eigenvalue $p \equiv F$ of the fermion charge operator, $\bar F = \psi^a \bar \psi^{a}$.
 Each component is normalized with the covariant measure,
 \begin{gather}
\label{measure}
 \mu   d^Dx   =  \sqrt {\det g}   d^Dx   =  \det h   d^d w d^d \bar w .
 \end{gather}
The supercharges (\ref{Qcovgen}) are conjugate to each other with respect to  this measure,
$\bar Q = \mu^{-1} Q^\dagger \mu$, where $Q^\dagger$ is a ``naive'' Hermitian conjugate.

It was shown in \cite{IvSm} that the supercharge $Q$ in (\ref{Qcovgen}) is isomorphic in this setting
to the exterior derivative operator $\partial$ and the Dolbeault complex is reproduced,
if choosing the function $G$ in a~special way,
 \begin{gather}
\label{G}
G   =  \frac 14 \ln \det h  .
 \end{gather}
 Another distinguished choice is $G = -(1/4) \ln \det h $ when the operator
$\bar Q$ is isomorphic to the antiholomorphic exterior derivative $\bar \partial$, and we arrive at the anti-Dolbeault
complex. For an arbitrary $G$, we are dealing with a {\it twisted} Dolbeault complex.

The Hamiltonian is given by the expression
  \begin{gather}
H  =  - \frac 12 \triangle^{\rm cov} + \ \frac 18 \left (R - \frac 12 h^{\bar k j}h^{\bar l t}h^{\bar i n}
 C_{j\,t\, \bar i}\,C_{\bar k\,\bar l\, n}\right) \nonumber\\
\phantom{H  =}{}  -  2 \langle \psi^a \bar \psi^{\bar b} \rangle\,e^k_a e^{\bar l}_{\bar b}\partial_k\partial_{\bar l} G
- \langle \psi^a \psi^c \bar \psi^{\bar b} \bar \psi^{\bar d} \rangle
 e^t_a e^j_c e^{\bar l}_{\bar b} e^{\bar k}_{\bar d}   (\partial_t\partial_{\bar l}{\,}h_{j\bar k}) .  \label{kvantH}
\end{gather}
  Here, $\langle \ldots \rangle$ denotes the Weyl-ordered products of fermions, $\langle \psi^a \bar \psi^{\bar b} \rangle =
 (\psi^a \bar \psi^{\bar b} - \bar\psi^{\bar b} \psi^a)/2 $, etc.
$R$~is the standard scalar curvature
of the metric $h_{j\bar k}$, while
\begin{gather}
\label{Cikl}
C_{j k\bar l} = \partial_{k} h_{j\bar l} - \partial_{j} h_{k\bar l}   , \qquad
C_{\bar j \bar k  l} =  (C_{j k \bar l})^* = \partial_{\bar k} h_{l \bar j} - \partial_{\bar j} h_{l \bar k}
\end{gather}
is the metric-dependent torsion tensor. The covariant Laplacian
 $\triangle^{\rm cov}$ is def\/ined with taking into account the torsion,
\begin{gather*}
-\triangle^{\rm cov}   = h^{\bar k j} \big( {\cal P}_j {\bar {\cal P}}_{\bar k} +
i \hat \Gamma^{\bar q}_{j \bar k} {\bar {\cal P}}_{\bar q}
+  {\bar {\cal P}}_{\bar k} {\cal P}_j + i \hat \Gamma^{s}_{\bar k j} {{\cal P}}_s \big),
 \end{gather*}
where ${\cal P}_j = \Pi_j + i \hat \Omega_{j, \bar b a} \langle\psi^a \bar \psi^{\bar b}  \rangle $ and
  ${\bar {\cal P}}_{\bar k} = \bar \Pi_{\bar k} -
i \hat {\bar \Omega}_{\bar k,  a \bar b} \langle \psi^a \bar \psi^{\bar b} \rangle  $
 with some particular torsionfull af\/f\/ine  and spin connections
(the so called Bismut connections
\cite{Bismut}),
 \begin{gather}
\hat{\Gamma}^M_{NK} = \Gamma^M_{NK} + \frac{1}{2}g^{ML}C_{LNK}, \label{genGamma}
\\
\hat \Omega_{M, AB}  = \Omega_{M, AB} + \frac{1}{2}e_A^K e_B^L C_{KML}, \qquad M \equiv \{m, \bar m\}. \label{hatOm}
\end{gather}

Note that this rather complicated expression for the Hamiltonian is greatly simplif\/ied in the K\"ahler
case. Then the  torsion (\ref{Cikl}) vanishes, the 4-fermion term in (\ref{kvantH}) vanishes too, and{\samepage
  \begin{gather*}
H_{\rm K\ddot{a}hl}   =    - \frac 12 \triangle^{\rm cov} + \frac R8
-  2 \langle \psi^a \bar \psi^{\bar b} \rangle e^k_a e^{\bar l}_{\bar b}\partial_k\partial_{\bar l} G ,
\end{gather*}
 where now $-\triangle^{\rm cov} = h^{\bar k j} \left( {\cal P}_j {\bar {\cal P}}_{\bar k} +
{\bar {\cal P}}_{\bar k} {\cal P}_j \right)$
with $\hat \Omega_{j, \bar b a} = \Omega_{j, \bar b a} =
 e^{\bar k}_{\bar b} \partial_j e^{\bar a}_{\bar k}$.}

In this paper, we are interested, however, with $S^4$, which is not K\"ahler and, as was mentioned,
not even globally complex.
This notwithstanding, one can write the supercharges~(\ref{Qcovgen})
and the Hamiltonian~(\ref{kvantH}) with the metric~(\ref{metrS4}), which is
well def\/ined everywhere on $S^4$ except the north pole and study the spectrum.

For sure, to determine the spectrum, we have to def\/ine f\/irst the {\it spectral problem} and to specify
the boundary conditions for the wave functions. There are two dif\/ferent reasonable choices:
$(i)$~We can consider the functions that are regular on $S^4$. $(ii)$~We can allow the singularity at the pole, but
require that the functions are square integrable with the measure~(\ref{measure}),
 \begin{gather*}
 \int   \frac {|\Psi|^2   d^2 w d^2 \bar w}{(1+\bar w w)^4}   <  \infty  .
  \end{gather*}
It turns out that, for the {\it first} spectral problem, the Hamiltonian is well def\/ined and Hermitian. However, the Hilbert
space of all nonsingular on $S^4$ functions does not constitute the domain of the supercharges: there exist nonsingular functions
$\Psi$ such that $ Q\Psi$ are singular. In physical language, this means that the supersymmetry is broken~-- some states do not
have superpartners. In mathematical language, this means that the Dolbeault complex is not well def\/ined on the manifolds
that are not complex, of which $S^4$ is an example.

What is, however, rather nontrivial and somewhat surprising is that,
in the Hilbert space of square integrable functions, everything works f\/ine. In the main body of the paper,
we will show that all excited square integrable states of the Hamiltonian are doubly degenerate (i.e.\ supersymmetry {\it is} there)
and that there are 3 bosonic zero modes such that the Witten index of this system is $I_W = 3$. In other words, even though
 the Dolbeault complex is not well def\/ined on $S^4$, there is a nontrivial
self-consistent way to def\/ine it on $S^4\backslash \{\cdot\}$.

There is a kinship between the problem under consideration and a problem of the Dirac complex on $S^2$ with noninteger
magnetic f\/lux~\cite{flux}. In both cases, the requirement for the spectrum to be supersymmetric  brings about
restrictions on the Hilbert space (see also~\cite{SSV}). However, for a noninteger f\/lux,
these restrictions are extremely stringent: they simply leave the Hilbert space empty,
a Dirac complex with noninteger f\/lux is not def\/ined. And, for $S^4$, the Hilbert
space of regular wave functions is not supersymmetric, while
its {\it extension}~-- the space of square integrable functions is.

\section{The Dolbeault Hamiltonian and its spectrum}\label{section2}

On $S^4$ with the metric (\ref{metrS4}) and with $G(\bar W, W)$ given by~(\ref{G}),
  the supercharges (\ref{Qcovgen}) acquire the following
simple form
\begin{gather}
 Q =i(1 + \bar w w) \psi_j \partial_j  + i \psi_j \psi_k
\bar\psi_j \bar w_k  , \nonumber \\
\bar Q  =  i \bar \psi_j \left[ (1 + \bar w w) \bar \partial_j  -
2w_j \right] + i \bar \psi_j \bar \psi_k  \psi_j  w_k  .\label{QS4}
 \end{gather}
The complicated expression (\ref{kvantH}) for the Hamiltonian also simplif\/ies a lot. There are three sectors:
$F=0$, $F=1$, and $F=2$.
Consider f\/irst the sector $F=0$. We obtain
 \begin{gather}
\label{HF0}
H^{F=0}   =  -(1 + \bar w w)^2 \partial_j \bar\partial_{j} + 2(1 + \bar w w) w_j \partial_j   .
\end{gather}
 It is instructive to compare this Dolbeault Laplacian with the standard covariant Laplacian on~$S^4$,
\begin{gather}
\label{lapl}
-\triangle_{S^4}   =   -(1 + \bar w w)^2 \bar \partial_j \partial_{j} +
(1 + \bar w w)
\big(w_j \partial_j  +  \bar w_j \bar \partial_j \big)  .
\end{gather}
Note that both (\ref{HF0}) and (\ref{lapl}) commute with the angular momentum operator
$m =  w_j  \partial_j  -  \bar w_j  \bar \partial_j $.\footnote{The Hamiltonian (\ref{HF0}) commutes
also with two other generators of $SU(2)$ such that the states represent $SU(2)$-multiplets.
The standard Laplacian (\ref{lapl}) has  $O(5)$ symmetry.}  The eigenvalues of $ m $ are integer.

The supercharges (\ref{QS4}) admit 3 normalizable zero modes satisfying
 $Q\Psi^{(0)} = \bar Q\Psi^{(0)} = 0$ in the sector
$F=0$,
 \begin{gather}
\label{zero}
 \Psi^{(0)}   =  1, \bar w_1, \bar w_2   .
 \end{gather}
They represent the ground states of the Hamiltonian (\ref{HF0}).

Consider now excited states.
 The eigenfunctions of the Hamiltonian can be sought for in the form
 \begin{gather}
\label{Ansatz}
 \Psi_{ms}   =  S_{ms}   F_{ms}(\bar w w )   ,
 \end{gather}
where $S_{ms}$ ($ m = 0, \pm 1, \ldots$; $ s = 0,1,\ldots$) are mutually orthogonal tensor structures
that vanish under the action of the ``naive Laplacian''
$\bar \partial_{j} \partial_j$. Each structure $S_{ms}$ has $2s + |m| + 1$ independent components
(and the corresponding energy level has degeneracy $2s + |m| + 1$). The explicit form of f\/irst few such structures is
   \begin{gather*}
 S_{00}  =  1 , \qquad S_{01}   =  w_j \bar w_k - \frac {\bar w w}2 \delta_{jk}   ,
\nonumber \\
S_{02}  =  w_i w_j \bar w_k \bar w_l - \frac {\bar w w}4 \left( w_i \bar w_k \delta_{jl} +
 w_i \bar w_l \delta_{jk} + w_j \bar w_k \delta_{il} + w_j \bar w_l \delta_{ik} \right)
+ \frac {(\bar w w)^2}{12} (\delta_{ik} \delta_{jl} + \delta_{il} \delta_{jk} )   , \nonumber \\
 S_{10}  =  w_j, \qquad  S_{11}   = w_i  w_j  \bar w_k - \frac {\bar w w}3 ( w_i \delta_{jk} +
 w_j \delta_{ik} )  , \qquad
  S_{-1,0}   =   \bar w_j   .
 \end{gather*}
 It is straightforward to see that the action of the Hamiltonian on the Ansatz~(\ref{Ansatz}) preserves its
tensor form. The radial dependence is then determined from the solution of scalar spectral equations for $F_{ms}$.
It is convenient to introduce the variable
\begin{gather*}
z   =  \frac {1-\bar w w}{1+\bar w w}
 \end{gather*}
(it is nothing but $\cos \theta$, $\theta$ being the polar angle on $S^4$).

The spectral equations acquire then the form
 \begin{gather*}
(z^2-1) F''(z) + 2(2z + m + 2s) F'(z) + \frac {4(m + s)}{1+z} F(z)  =  \lambda F(z)  , \qquad m \geq 0   ,\nonumber \\
 (z^2-1) F''(z) + 2(2z + |m| + 2s) F'(z) + \frac {4s}{1+z} F(z)   =  \lambda F(z)  , \qquad m \leq 0 .
 \end{gather*}
     Their formal solutions are
\begin{gather}
F(z)  = (1 + z)^{\gamma_{ms}} P_n^{|m|+ 2s +1, \pm \Delta_{ms}}(z) ,
\nonumber \\
\lambda_{msn} = \gamma_{ms}^2 + 3\gamma_{ms} + n(n + |m| + 2s
+ 2 \pm \Delta_{ms} ) ,\label{solFz}
 \end{gather}
with
  \begin{gather}
\label{gam}
 \gamma_{ms}   =  \frac {|m| + 2s -1 \pm \Delta_{ms} }2
  \end{gather}
 and
 \begin{gather*}
 \Delta_{ms}  =  \sqrt{(1-m - 2s)^2 + 8(m + s)} , \qquad m \geq 0   , \nonumber \\
\Delta_{ms}   =  \sqrt{(1-|m| - 2s)^2 + 8s}   , \qquad m \leq 0  .
 \end{gather*}
$P_n^{\alpha, \beta}$ ($n = 0,1,\ldots$) are the Jacobi polynomials,
\begin{gather*}
P^{\alpha,\beta}_n(z)   =  \frac 1{2^n} \sum_{k=0}^n   \begin{pmatrix} n+\alpha \\ k \end{pmatrix}
  \begin{pmatrix} n+\beta \\ n- k \end{pmatrix} (1 + z)^k (z-1)^{n-k}   .
 \end{gather*}
For $\alpha > -1$, $\beta > -1$, the Jacobi polynomials are mutually orthogonal
on the interval $z \in (-1,1)$ with the weight
$\mu = (1-z)^\alpha (1+z)^\beta$.

Not all the solutions in (\ref{solFz}) are admissible, however. One can observe the following:
 \begin{itemize}\itemsep=0pt
\item  First of all,  all the solutions with  $s > 0$ and/or $m > 0$ and the negative
sign of $\Delta_{ms}$ in~(\ref{solFz}),~(\ref{gam})  are not square integrable and should not be included in
the spectrum.
If $s=0$ and $m \leq 0$, the solution with negative sign of  $\Delta_{ms}$ are not independent being expressed into
the solutions with positive sign in virtue of the identity
  \begin{gather}
\label{relJacobi}
P^{\alpha, -\beta}_{n+\beta}(z) = 2^{-\beta} (z+1)^\beta \frac {n!
(n+\alpha + \beta)!}{(n+\alpha)! (n+\beta)!} P_n^{\alpha, \beta}(z)   ,
 \end{gather}
which holds for integer $\alpha$~\cite{WuYang}.
\item On the other hand, the solutions with positive sign of $\Delta_{ms}$
and with  nonnegative $m$  are all not only square integrable,
but also nonsingular on $S^4$. In addition, they belong to the domain of $Q$:
$ Q \Psi_{m \geq 0, s}$ is never singular.
 \item Most of the solutions with  $m < 0 $ also have this property. However, there are three distinguished families of
solutions: the solutions
 \begin{gather}
\label{m=-1}
 \Psi_{-1,0,n}   =  \bar w_j P_n^{2,0}(z)  ,
 \end{gather}
 the solutions
 \begin{gather}
\label{m=-2}
 \Psi_{-2,0,n}   =  \frac {\bar w_j \bar w_k}{1 + \bar w w} P_n^{3,1}(z)   ,
 \end{gather}
and the solutions
\begin{gather}
\label{m=-3}
 \Psi_{-3,0,n}   =  \frac {\bar w_j \bar w_k \bar w_l}{(1 + \bar w w)^2 }
 P_n^{4,2}(z)  .
 \end{gather}

$(i)$ The functions (\ref{m=-1}) are all singular at inf\/inity, but integrable. Two lowest such functions $\Psi = \bar w_j$
are zero modes of the Hamiltonian (\ref{HF0}). The functions ${ { Q}} \Psi_{-1,0,n}$ are less singular: they do not grow at inf\/inity
(though do not have a def\/inite value there when $n > 0$).

$(ii)$ The functions (\ref{m=-2}) are bounded at inf\/inity. The supercharge action produces growing functions,
${{ Q}} \Psi_{-2,0,n} (w=\infty) = \infty$. Still, ${{ Q}} \Psi_{-2,0,n}$ is square integrable.

$(iii)$ The functions (\ref{m=-3}) are regular at inf\/inity. The supercharge action produces  singular bounded functions.

\item  Note that  if $\Psi$ is a
{\it non-normalizable} eigenfunction in the sector $F=0$, the function $Q \Psi$ is also not
normalizable. Indeed, the action of $ Q$ brings about
generically an extra power of $|w|$, which makes the divergence still stronger.
 An exception would  only be provided by the
functions with the asymptotics $\propto \bar w_{j_1} \cdots \bar w_{j_k}$  at inf\/inity. But the only {\it eigenfunctions}
 with
such asymptotics are written in equation~(\ref{m=-1}). They are normalizable.
 \end{itemize}

Consider now the sector $F=2$. The Hamiltonian is
\begin{gather*}
H^{F=2} \ =\ -(1 + \bar w w)^2 \bar \partial_j \partial_j + 2(1 + \bar w w) w_j \partial_j
+ 2(2 + \bar w w)   .
\end{gather*}
 The eigenfunctions  have the same form as in (\ref{Ansatz}), (\ref{solFz}), (\ref{gam})   with
  modif\/ied
\begin{gather*}
 \Delta_{ms}^{F=2}   =  \sqrt{(1- m - 2s)^2 + 8(m+s+1)} , \qquad m \geq 0   , \nonumber \\
 \Delta_{ms}^{F=2}  = \sqrt{(1-|m| - 2s)^2 + 8( s+1)} , \qquad  m \leq 0   .
 \end{gather*}
Again, almost all functions with negative sign in (\ref{gam}) are not normalizable. The exceptions are the sectors
$s=0$, $m = 0, -2$, where the functions with the negative sign are expressed into the functions with positive sign.
 Speaking of the latter, they are not only normalizable, but also nonsingular in this case.
 All these functions are annihilated by $Q$ and belong
to the domain of~${ {\bar Q}}$,
${{\bar Q}} \Psi^{F=2}$ being regular on~$S^4$.

In the sector $F=1$, the wave functions have two components,
$\Psi^{F=1}  = \psi_j C_j(\bar w_k, w_k)$. No new zero modes appear. Indeed,
if the Hamiltonian
{\it had} zero modes in this sector, they would
satisfy the conditions $Q \Psi = { {\bar Q}} \Psi = 0$ giving
 \begin{gather*}
 (1 + \bar w w) \partial_{[j} C_{k]} - \bar w_{[j}  C_{k]}  =  0   , \nonumber \\
 (1 + \bar w w) \bar \partial_k C_k - 3 w_k C_k  =  0   .
 \end{gather*}
The f\/irst equation can be rewritten as
\[
 \partial_{[j} \left[ \frac {C_{k]}}{1+\bar w w} \right]   =  0
 \]
with a generic solution $C_k = (1+ \bar w w) \partial_k \Phi$. Then the second equation gives
$H^{F=0} \Phi = 0$. If $C_k$ is normalizable, $\Phi$ must also be normalizable (modulo a pure antiholomorphic
part). But we have seen that the only normalizable zero modes of $H^{F=0}$ are 1 and $\bar w_j$
annihilated by holomorphic derivatives and giving $C_k = 0$.\footnote{Note that if one lifts the normalizability condition, a nontrivial solution of the equation
$H^{F=0} \Phi = 0$ exists: $\Phi = \bar w w + 2 \ln (\bar w w) - \frac 1{\bar w w}$ giving
 $C_k  = \bar w_k (1 + \bar w w)^3/(\bar w w)^2$.}

To f\/ind the nonzero modes in the sector $F=1$, one needs not to solve the Schr\"odinger equation again. All such normalizable
functions are obtained by the action of $Q$ or ${ {\bar Q}}$ onto the normalizable functions in the sectors $F=0$
or $F=2$, correspondingly. This follows from the last itemized statement above, which is valid also in the sector $F=2$.
 By construction, these functions are annihilated by $Q$ or $\bar Q$ and belong to the domain
of $\bar Q$ or $Q$, correspondingly.

\subsection{Twisted Dolbeault complex}

The Hamiltonian (\ref{kvantH}) is supersymmetric not only under the condition
(\ref{G}) that distinguishes the pure Dolbeault complex, but also with other choices of $G$ describing
 twisted Dolbeault complexes.

 First of all, we can set $G=0$. As was shown in \cite{IvSm},
the Hamiltonian (\ref{kvantH}) with $G=0$ coincides  with the extended $N=4$ supersymmetric Hamiltonian
written in~\cite{Konush},
 \begin{gather}
\label{HamKon}
H   =  - \frac 12 f^3 \partial_M^2 \frac 1f - \frac 12\, \psi
\sigma_{[M}^\dagger \sigma_{N]} \bar \psi   f (\partial_M f) \partial_N + f (\partial^2 f)
\left ( \psi \bar \psi - \frac 12 (\psi \bar \psi)^2 \right)  ,
 \end{gather}
with  $f = 1 + x_M^2/2$.

This model belongs to the class of the so called ``hyperk\"ahler with torsion'' (HKT) mo\-dels~\cite{Howe},
which were  classif\/ied using the harmonic superspace formalism in recent \cite{Delduc}.
The Hamiltonian~(\ref{HamKon}) does not admit normalizable zero-energy solutions and  its index is zero.

Consider now a model with
 \begin{gather*}
 G   =  \frac q4 \ln \det h = -q \ln(1 + \bar w w)
 \end{gather*}
with an integer $q >1$. The supercharges are then
  \begin{gather*}
 Q  =  i(1 + \bar w w) \psi_j \partial_j  + i (q-1)  \bar w w  \psi_j \bar w_j
+
i \psi_j \psi_k \bar\psi_j  \bar w_k   , \nonumber \\
\bar Q  =  i \bar \psi_j \left[ (1 + \bar w w) \bar \partial_j  -
(q+1) w_j \right] + i \bar \psi_j \bar \psi_k  \psi_j  w_k  .
 \end{gather*}
 The zero modes all dwell in the sector $F=0$. They have the form
 \begin{gather*}
\Psi^{(0)}  = \frac {P(\bar w)}{(1+\bar w w)^{q-1}}  ,
 \end{gather*}
where $P(\bar w)$ is an antiholomorphic polynomial of degree $2q-1$.
It has $2q^2 + q$ independent coef\/f\/icients,
which gives  $2q^2 + q$ independent zero modes.

When $q$ is negative, the analysis is similar.
It gives $2q^2-q$ zero modes in the sector $F=2$. The same consideration
 as in the
pure Dolbeault case displays the absence of the normalized
zero modes in the sector $F=1$.
 The f\/inal result for the index of the twisted Dolbeault complex~is
 \begin{gather}
\label{Indexq}
 I(q)  = 2q^2 + |q|  .
 \end{gather}

\section{The index and the functional integral}\label{section3}

As was mentioned, for pure Dolbeault complex, there are 3 bosonic zero modes (\ref{zero})
in the sector $F=0$ and no zero modes in the
other sectors. This means
that the Witten index of this system,
\begin{gather*}
I_W  =  {\rm Tr} \big\{(-1)^F e^{-\beta H} \big\}
 \end{gather*}
 is equal to 3.

For {\it compact} complex manifolds, the Witten index of the supersymmetric Hamiltonian (\ref{kvantH}) under
the condition (\ref{G}) is known to mathematicians by the name of {\it arithmetic genus} of the manifold. This invariant
admits an integral representation known as the Hirzebruch--Riemann--Roch theorem \cite{Hirz},\footnote{Following the ideas of \cite{A-GFW}, it has been recently derived also in physical way
by studying the path integral for the supersymmetric partition function~\cite{HRR}.}
 \begin{gather*}
I \ =\ \int  \, {\rm Td}(TM) ,
 \end{gather*}
where the symbol Td$(TM)$ ({\it Todd class of a complex tangent bundle} associated with the manifold~$M$) is spelled out as
 \begin{gather}
\label{Todd}
 {\rm Td}(TM)   =
  \prod_{\alpha = 1}^n \frac {\lambda_\alpha/2\pi}{1 - e^{-\lambda_\alpha/2\pi} }   ,
  \end{gather}
where $\lambda_\alpha$ are eigenvalues of the curvature matrix corresponding to this bundle\footnote{For (\ref{Todd}) to be correct and simply to make sense, the connection and its curvature should respect the complex
structure. For example, the Bismut connection (\ref{genGamma}) is appropriate for this purpose, while the usual
torsionless Levi-Civita connection is not, if the manifold is not K\"ahler.}.

The representation (\ref{Todd}) can be derived by using the fact that the sum of the supercharges~(\ref{QS4}) can be interpreted
as a Dirac operator involving an Abelian  gauge f\/ield and torsions. (The presence of torsions is a complication
that distinguishes the HRR theorem from a version of the Atiyah--Singer theorem
discussed usually by physicists.) It is important in this derivation that the  gauge f\/ield
represents a regular f\/iber bundle on the manifold, while torsions are regular tensors.

In our $ S^4$ case, however, these
conditions are not fulf\/illed: the gauge f\/ield and the torsion are singular at $w = \infty$.
Indeed, the torsion (\ref{Cikl}) with the metric (\ref{metrS4}) behaves at inf\/inity as $\sim |x|^{-5}$.
Then $g^{MN} g^{PQ} g^{ST} C_{MPS} C_{NQT} \sim (x^{4})^3 \cdot (x^{-5})^2 \sim x^2$ and diverges.
The gauge f\/ield (\ref{A}), (\ref{G}) is rather peculiar. It is {\it disguised} as a benign f\/iber bundle  having
 an integer Chern class
 \begin{gather}
\label{Chern}
 {\rm Ch}_2 =  \frac 1{8\pi^2} \int F \wedge F   =   2   .
 \end{gather}
However, a  topologically nontrivial $U(1)$ bundle on $S^4$ {\it does} not exist because $\pi_3[U(1)] = 1$.\footnote{On the other hand, topologically
nontrivial bundles on $S^4$ with non-Abelian gauge groups ({\it instantons}), of course, exist.} Indeed, the f\/ield strength tensor
is singular in this case, which manifests itself in the fact that the ``action integral'' $\sim
\int d^4 x \sqrt{g}   F_{MN} F^{MN}$  diverges logarithmically.

For such a singular Dirac operator, one cannot get rid of torsions by a smooth deformation
(a key step in the derivation of~(\ref{Todd})), because the index integral may in this case acquire contributions from total
derivatives of singular expressions. Moreover, with all probability, the Dirac operator on $S^4$ with the singular
f\/ield~(\ref{A}) and
{\it without} torsions does not describe a benign supersymmetric system~-- the situation must be the same as for
the gauge f\/ield on~$S^2$ with non-integer magnetic f\/lux~\cite{flux}.

Having no further mathematical methods at our disposal (at least, we are not aware of such methods), we can try to
calculate the Witten index in  a  physical way by  evaluating directly
 the corresponding path integral,
  \begin{gather*}
 I =   \int d\mu \exp \left\{ - \int_0^\beta L_E(\tau) d\tau \right \}  ,
\end{gather*}
 where $L_E$ is the Euclidean Lagrangian of our supersymmetric quantum system, $d\mu$ is the appropriate
functional integral measure, and the periodic boundary conditions are imposed onto all variables.

 For most supersymmetric systems, this integral is reduced for small $\beta$ to an ordinary phase space integral \cite{Cecotti}.
This is true e.g.\ for a supersymmetric Hamiltonian describing the de Rham complex on a compact manifold,
where the Witten
index is given by its Euler characteristics. For the Dirac complex on compact
manifolds,  the situation is more complicated, a naive semiclassical reduction is not
justif\/ied and one has to perform
a honest calculation of the path integral in  the one-loop approximation~\cite{A-GFW}, which is not so trivial
 (see~\cite{IvSm} for detailed pedagogical explanations). For the Dolbeault complex on compact non-K\"ahler
complex manifolds, the life is still more dif\/f\/icult.
Generically, one has to perform a~two-loop calculation for $4d$ and $6d$ manifolds, a~three-loop
calculation for $8d$ and $10d$ manifolds, etc.  This complication is due to the appearance of the new 4-fermionic term
in the Lagrangian,
 \begin{gather}
L_E   =  \frac{1}{2}\left[ g_{MN} \dot x{\,}^M \dot x{\,}^N + g_{MN}\,\psi^M \hat{\nabla} \psi^N
+ \frac{1}6  \partial_P C_{MNT} \psi^P\psi^M\psi^N\psi^T \right] \nonumber\\
\phantom{L_E   =}{} -  i A_M \dot{x}^M  + \frac i2 F_{MN} \psi^M \psi^N   \label{LE}  
\end{gather}
($\hat{\nabla} \psi^M = \dot{\psi}^M + \hat \Gamma^M_{NK} \dot{x}^N \psi^K$  is the Bismut covariant derivative).
For example, for a 4-dimensional manifold, the leading (at small $\beta$) contribution to the index is
 \begin{gather}
\label{Ising}
I   \sim   \frac 1\beta \int   d^4 x\,  \epsilon^{MNPQ} \partial_M C_{NPQ}  .
 \end{gather}
For sure, the integrand  is a total derivative here and the integral vanishes, but the appearance of the large factor $1/\beta$
does not allow one  to ignore 2-loop corrections anymore. They are essential. For $8d$ manifolds, the leading contribution in
the integrand is of order $\sim 1 /\beta^2$, and this makes essential 3 loop contributions, etc.

As was mentioned above, for compact manifolds, one needs not actually to come to grips with these complicated
multiloop contributions. One can, instead, perform a smooth deformation that kills the torsion and makes the problem and the
corresponding path integral tractable. For a particular class of manifolds  where the fermion term
in~(\ref{LE}) vanishes (the so called SKT manifolds), this program was in fact carried out  in~\cite{Bismut}. (In this mathematical
paper, path integrals and
supersymmetry were not mentioned and the author described the results in the language of {\it heat kernel} technique,
 which is equivalent, however, to the path integral approach.) The generic case is discussed in~\cite{HRR}.

For $S^4$, we have no other choice than to try to evaluate the path integral directly. In 4~dimensions, this is dif\/f\/icult,
but feasible and, in the case when the Dirac operator involves only torsions, but no extra gauge f\/ield ($G=0$ in our language),
has been performed in \cite{Waldron}. The index integral has been represented in these papers as
   \begin{gather}
I   =  - \frac 1{4\pi^2} \int d^4x \, \sqrt{g} \bigg\{ \frac 1{2\beta} \nabla_M B^M     \nonumber \\
\phantom{I   =}{}+\frac 1{192} \frac 1{\sqrt{g}}
\epsilon^{RSKL} \left[ R_{MNRS} R^{MN}_{\ \ \ \ KL} + \frac 12 B_{RS} B_{KL} \right]
  + \frac 1{24} \nabla_M {\cal K}^M
\bigg\} + {\cal O} (\beta)\label{IWald}
 \end{gather}
 with
 \begin{gather}
\label{KB}
{\cal K}^M = \left( \nabla^N \nabla_N + \frac 14 B^N B_N +
\frac 12 R \right) B^M, \qquad B_{MN} = \nabla_M B_N - \nabla_N B_M  .
 \end{gather}
Here  $\nabla_M$ is the standard Levi-Civita covariant derivative and $R_{MNRS}$ is the standard Riemann tensor. $B_M$ is the
axial vector dual to the torsion tensor,
 \begin{gather*}
B^M = \frac 1 {6\sqrt{g}}  \epsilon^{MNPQ} C_{NPQ}  =   2x^M \big(1 + x^2/2\big)  .
 \end{gather*}
The f\/irst term in equation~(\ref{IWald}) is a singular ($\sim 1/\beta$ )
 but vanishing
integral of a total derivative. It was discussed before. The last term also represents a total derivative and also vanishes\footnote{This vanishing is due to a certain cancellation. One can easily check that {\it individual} terms
$\sim \nabla^N \nabla_N$ and $\sim B^N B_N $ in (\ref{KB})
are expressed into the integral $\sim \int d^4x \, \partial_M (x^M/x^4)$ that does not vanish.
These contributions cancel, however, in the sum.}.  The ``f\/ield strength'' $B_{MN}$ is zero in our case. The quadratic in the
Riemann tensor integral is proportional to a certain topological invariant called Hirzebruch signature. For $S^4$, it vanishes.

As a result, the index of the corresponding supersymmetric Hamiltonian vanishes.
This agrees with the direct analysis of its spectrum (see the remark after equation~(\ref{HamKon})).

When $G \neq 0$, the Lagrangian and the Hamiltonian involve an extra gauge f\/ield. The functional integral
for the index acquires the (tree-level) contribution~(\ref{Chern}).

On top of this, there might have been a 1-loop contribution associated with the 4-fermion term.
To evaluate it, it is convenient to expand the periodic f\/ields $x^M(\tau)$ and $\psi^M(\tau)$
into the Fourrier modes,
 \begin{gather*}
x^M (\tau) =  x^{M}_0 + \sum_{m \neq 0} x^{M}_m e^{2\pi i m \tau/\beta},\qquad
\psi^M (\tau) =  \psi^{M}_0 + \sum_{m \neq 0} \psi^{M}_m e^{2\pi i m \tau/\beta}  ,
 \end{gather*}
with integer $m$ ($\bar x_m^M = x_{-m}^M$, $\bar \psi_m^M = \psi_{-m}^M$). We obtain instead of (\ref{Ising})
 \begin{gather}
I   \sim   \frac 1\beta \int   d^4 x_0\,  \epsilon^{MNPQ} \partial_M C_{NPQ}   \mu\nonumber\\
\phantom{I   \sim }{}
\times \prod_{M, m \neq 0} dx^M_m d\psi^M_m
\exp \Bigg\{ {-} \frac  1{2\beta} \sum_m (2\pi m)^2 x^M_m x^N_{-m}
 \left( g_{MN}  - \frac{\beta F_{MN}} {2\pi m}  \right)\nonumber \\
\phantom{I   \sim }{}
  +i \sum_m (2\pi m)   \psi^M_m \psi^N_{-m}  \left( g_{MN}  - \frac{\beta F_{MN}} {2\pi m}  \right)   \Bigg\}  ,\label{Isinggauge}
 \end{gather}
where $\partial C$, $g$, $F$, and the inf\/inite factor $\mu$\,\footnote{It can be made f\/inite when imposing an ultraviolet cutof\/f,
see  equation~(6.27) of~\cite{IvSm}. When $F=0$,
\[
\mu \int   \prod_{M, m \neq 0} dx^M_m d\psi^M_m
\exp \{ \cdots \}   =  1   .
\]}
depend on the zero coordinate modes $x_0^M$.

The individual contributions due to bosonic and fermionic Gaussian integrals are nontrivial.
For example, the fermionic integral gives
 \begin{gather*}
 {\rm fermion\ factor}   =    \prod_{ m = 1}^\infty  \det \left\| \delta_M^{\ N} - \frac {\beta F_M^{\ N}}{2\pi m} \right \|
=   \det\nolimits^{ 1/2}   \frac {2\sin (\beta F/2 ) }{\beta F  } .
 \end{gather*}
 Formally, the correction to unity is proportional to $\beta^2$, but it multiplies
${\rm Tr}\{ F^2\} = - F_{MN} F^{MN}$, which grows at inf\/inity $\sim x^4$.  The integral is then
saturated by large $x$ values, and, as a result, the correction
is of order $\beta$. When multiplied by the overall factor~$1/\beta$ in front of the integral, this gives a
correction of order 1 to the index.

Anyway, as is clear from (\ref{Isinggauge}), this  fermionic correction exactly cancels the bosonic one
(obviously, this cancelation is due to supersymmetry), and we have to conclude that the functional
integral calculation gives the value (\ref{Chern}) for the index, which contradicts the direct analysis above
giving $I = 3$. In addition, for the twisted Dolbeault complex, we obtain
 \begin{gather*}
I_{\rm funct.\ int.}   =  2q^2   ,
 \end{gather*}
which contradicts the estimate (\ref{Indexq}) above.

Certainly, this mismatch is disappointing and paradoxical. We want to emphasize, however, that there is
no {\it logical} contradiction here. We calculated the functional integral by semiclassical methods expanding
it in $\beta$. In particular, we studied only one-loop corrections to the index associated with gauge f\/ield, because
two- and higher-loop corrections are suppressed by the naive~$\beta$ counting.
 We have seen, however, that this expansion breaks down near the singularity where~$\beta$ is multiplied
by a large factor~$\sim x^2$. In this situation, one cannot reliably justify ignoring higher-loop contributions.
They can give something (though we do not see at the moment how this can come about).

Note that there are some other examples where the presence of singularities invalidates the
 semiclassical calculation of the path integral. In particular, in~\cite{Blok}, we constructed SQM systems
associated with chiral supersymmetric gauge theories in f\/inite volume. The Hamiltonian
of these systems is singular near the origin, $H \sim 1/x^2$. And, though this singularity is of repulsive benign
nature, unitarity is not broken, and the spectrum of the Hamiltonian is discrete, the semiclassical approximation
for the path integral breaks down near the origin. This manifests itself in the senseless
fractional values of the path integral for the index evaluated at the leading order.

Def\/initely, more studies of this very interesting question are necessary.

\section[$S^6$]{$\boldsymbol{S^6}$}\label{section4}

A similar analysis can be done for $S^6$ and also for higher even-dimensional spheres.
The metric of $S^6$ is still given by equation~(\ref{metrS4}) where now $j = 1,2,3$. The supercharges
of the SQM system describing
the pure Dolbeault complex have almost
the same form as for $S^4$,
\begin{gather*}
 Q  =  i(1 + \bar w w) \psi_j \partial_j  + i \psi_j \psi_k
\bar\psi_j \bar w_k   , \nonumber \\
\bar Q = i \bar \psi_j \left[ (1 + \bar w w) \bar \partial_j  -
3w_j \right] + i \bar \psi_j \bar \psi_k  \psi_j  w_k   .
 \end{gather*}
The Hamiltonian in the sector $F=0$ is
\begin{gather*}
H^{F=0}   =  -(1 + \bar w w)^2 \partial_j \bar\partial_{j} + 4(1 + \bar w w) w_j
\partial_j   ,\qquad j=1,2,3 ,
\end{gather*}
to be compared with the standard covariant Laplacian on $S^6$,
\begin{gather*}
-\triangle_{S^6}  =  -(1 + \bar w w)^2 \bar \partial_j \partial_{j} +
2(1 + \bar w w)
(w_j \partial_j  +  \bar w_j \bar \partial_j ) .
\end{gather*}
We choose the basis
 \begin{gather*}
\label{AnsatzS6}
 \Psi_{pq}   =  T_{pq}   F(\bar w w )   ,
 \end{gather*}
where a tensor structure $T_{pq}$ having $p$ factors  $w$ and $q$ factors
 $\bar w$ and annihilated by $\partial_j \bar \partial_j $
represents a $\begin{pmatrix} p \\ q \end{pmatrix}$ multiplet of $SU(3)$
and has $(p+1)(q+1)(p+q+2)/2$ independent components. For example,
 \begin{gather*}
T_{11}   =   w_j \bar w_k - \frac {\bar w w}3 \delta_{jk}
 \end{gather*}
is an octet\footnote{The $(p,q)$ notation can also be used for $S^4$, in which case $m = p-q$ and $s = \min\{ p,q\}$.}.
 The spectral equations for the coef\/f\/icients $F(z)$ are
 \begin{gather*}
\big(z^2-1\big) F''(z) + 2(3z + p + q) F'(z) + \frac {4p}{1+z} F(z)  =  \lambda F(z)  .
 \end{gather*}
Their solutions are
 \begin{gather}
\label{solFz3}
F(z)   =  (1 + z)^{\gamma_{pq}} P_n^{p+q+2, \pm \Delta_{pq}}(z)  ,
\qquad  \lambda_{pqn} =  \gamma_{pq}^2 + 5\gamma_{pq} + n(n + p+q
+ 3 \pm \Delta_{pq} )   ,
 \end{gather}
with
\begin{gather}
\label{gam3}
 \gamma_{pq}   =  \frac {p+q -2 \pm \Delta_{pq} }2
  \end{gather}
 and
 \begin{gather*}
 \Delta_{pq}  =  \sqrt{(2-p-q)^2 + 16p}   .
 \end{gather*}
 The observations to be made are exactly parallel to the observations in the $S^4$ case.
In particular,
\begin{itemize}\itemsep=0pt
\item  The solutions with  $p > 0$  and the negative
sign of $\Delta_{pq}$   are not square integrable and should not be included in
the spectrum.
If $p=0$ or $p = q = 1$, the solutions with negative sign of  $\Delta_{ms}$ are not independent being expressed into
the solutions with positive sign in virtue of~(\ref{relJacobi}).
\item On the other hand, the solutions with positive sign of $\Delta_{pq}$
and $p > 0$  are all not only square integrable,
but also nonsingular on $S^6$. In addition, they belong to the domain of~$Q$
(meaning here that $ Q \Psi$ are all normalizable).
 \item The solutions with  $p= 0$ and $q > 4$ also have this property.
\item
The normalizable\footnote{This means here that
\[
 \int   \frac {|\Psi|^2 \, d^3 w d^3 \bar w}{(1+\bar w w)^6}   <  \infty  ,
\]
such that the singularity $\Psi \sim |w|^2$ is still allowed, while $\Psi \sim |w|^3$ is already not.}
 families of solutions with $p=0$ and $q = 0,1,2,3,4,5$ are
 \begin{gather*}
 \Psi_{00n}  =   P_n^{22}(z), \qquad  \Psi_{01n} = \bar w_j   P_n^{31}(z)  , \nonumber \\
 \Psi_{02n}  =  \bar w_j \bar w_k \, P_n^{40}(z)   , \qquad
\Psi_{03n} =   \frac {\bar w_j \bar w_k \bar w_l}{1 + \bar w w} \, P_n^{51}(z)  , \nonumber \\
\Psi_{04n}  =  \frac {\bar w_j \bar w_k \bar w_l \bar w_p}{(1 + \bar w w)^2} P_n^{62}(z) , \qquad
\Psi_{05n} = \frac {\bar w_j \bar w_k \bar w_l \bar w_p \bar w_s} {(1 + \bar w w)^3} P_n^{73}(z) .
 \end{gather*}
The families with $q=1,2,3$ grow at inf\/inity. The functions $\Psi_{04n}$ are bounded, but still singular
($\Psi(\infty)$ is not def\/ined). However,
one cannot restrict oneself with the regular functions. The last family in the list above
is regular on $S^6$, but would not belong to the domain of $Q$ in this case: $Q \Psi_{05n}$ is not regular
at inf\/inity.
In addition, by the same token as for $S^4$,
the family
$ Q   \Psi_{01n}$ is regular on $S^6$, but does not belong to the domain of $\bar Q$ (because
$\Psi_{01n}$ are singular).
 \end{itemize}
For normalizable functions, we probably have a nice complex.
As we have just shown, all normalizable functions in the sector $F=0$ have normalizable superpartners.

The Hamiltonian in the sector $F=3$ is
\begin{gather*}
H   =  -(1 + \bar w w)^2 \partial_j \bar \partial_j + (1+\bar w w)\left(
\bar w_j \bar \partial_j + 3w_j \partial_j \right) + 3(3+\bar w w)  .
 \end{gather*}
 The spectral equations for the coef\/f\/icients $F(z)$ of the structures
$T_{pq}$ are
 \begin{gather*}
\big(z^2-1\big) F''(z) + 2(3z + p + q) F'(z) + \frac {3(p+1) + q}{1+z} F(z)  =  (\lambda - 6) F(z) ,
 \end{gather*}
and the solutions are also given by (\ref{solFz3}), (\ref{gam3}) with
\begin{gather*}
 \Delta_{pq}^{F=3}   =    \sqrt{(2-p-q)^2 + 4[3(p + 1) + q] }   .
 \end{gather*}
The eigenfunctions have better convergence here than in the sector $F=0$. Actually, all normalizable
eigenfunctions as well as their superpartners  (they have fermion charge $F=2$) are regular on $S^6$.

 To {\it prove} that the Dolbeault complex is well def\/ined in this case in the space of square integrable
functions, we have also to solve the Schr\"odinger equation
 in the sectors $F=1$ and $F=2$. In this case, it is more dif\/f\/icult than for $S^4$ because {\it some} states
in the sector $F=1$ are annihilated by $\bar Q$ and cannot be found as superpartners of the states
in the sector $F=0$. Likewise, there are states in the sector $F=2$ that are not superpartners to the states
with $F=3$. (These new states are superpartners to each other.) A special analysis of the matrix
Schr\"odinger equation is thus required. We do not think,
however, that such an analysis would unravel unpleasant surprises and believe that the Dolbeault
complex is well def\/ined on~$S^6\backslash \{\cdot\} $.

The Witten index of this system is equal to
 \begin{gather*}
I^{S^6\backslash \{\cdot\}}   =  1 + 3 + 6 = 10 .
 \end{gather*}
 (There is one zero mode, $\Psi = 1$,
in the sector $F=0$, $p=q=0$, three zero modes, $\Psi = \bar w_j$, in the sector $F=0$, $p = 0$, $q = 1$ and six
zero modes $\Psi = \bar w_j \bar w_k $, in the sector $F=0$, $p = 0$, $q = 2$.
No zero modes in the other sectors are present.)
Generalizing this analysis to higher spheres, we obtain the result
 \begin{gather*}
  I^{S^{2d}\backslash \{\cdot \}}   =  C^{d-1}_{2d-1}
\end{gather*}
   for the index of the pure Dolbeault complex.

Again, we can try to make contact of this result with functional integral calculations.
Unfortunately, this does not work better here than in the $S^4$ case. At the tree level, one obtains
 a~{\it fractional} contribution to the path integral,
\begin{gather*}
  \frac 1{48\pi^3} \int F \wedge F \wedge F  =   \frac 92  .
 \end{gather*}
 The one-loop contribution associated with the gauge f\/ield
vanishes by the same token as for~$S^4$ (see equation~(\ref{Isinggauge})
and the discussion thereabout).
Higher loops seem to be suppressed for small~$\beta$, but the presence of singularity does not allow  one to make
a def\/inite statement \dots.

 \subsection*{Acknowledgements}

I am indebted to G.~Carron, E.~Ivanov, and V.~Roubtsov for useful discussions.

\pdfbookmark[1]{References}{ref}
\LastPageEnding

\end{document}